\documentclass[aps,reprint,superscriptaddress,showpacs]{revtex4-1}
\usepackage{graphicx}
\usepackage{dcolumn}
\usepackage{bm}
\begin{document}
\title{Large Dispersive Shift of Cavity Resonance Induced by a Superconducting Flux Qubit in the Straddling Regime}

\author{K. Inomata} \email[k-inomata@zp.jp.nec.com]{}
\affiliation{RIKEN Advanced Science Institute, Wako, Saitama 351-0198, Japan}

\author{T. Yamamoto}
\affiliation{RIKEN Advanced Science Institute, Wako, Saitama 351-0198, Japan}
\affiliation{NEC Smart Energy Research Laboratories, Tsukuba, Ibaraki 305-8501, Japan}

\author{P.-M. Billangeon}
\affiliation{RIKEN Advanced Science Institute, Wako, Saitama 351-0198, Japan}

\author{Y. Nakamura}
\affiliation{RIKEN Advanced Science Institute, Wako, Saitama 351-0198, Japan}
\affiliation{Research Center for Advanced Science and Technology,
The University of Tokyo, Meguro-ku, Tokyo 153-8904, Japan}
 
\author{J. S. Tsai}
\affiliation{RIKEN Advanced Science Institute, Wako, Saitama 351-0198, Japan}
\affiliation{NEC Smart Energy Research Laboratories, Tsukuba, Ibaraki 305-8501, Japan}

\date{\today}

\begin{abstract}
We demonstrate enhancement of the dispersive frequency shift in a coplanar waveguide resonator induced by a capacitively-coupled superconducting flux qubit in the straddling regime.
The magnitude of the observed shift, 80~MHz for the qubit-resonator detuning 
of 5~GHz, is quantitatively explained by the generalized Rabi model which takes into account the contribution of the qubit higher energy levels. 
By applying the enhanced dispersive shift to the qubit readout, we achieved 90$\%$ contrast of the Rabi oscillations 
which is mainly limited by the energy relaxation of the qubit. 
\end{abstract}


\maketitle
High-fidelity readout is essential for quantum information processing. 
In the architectures of circuit quantum electrodynamics based on 
superconducting circuits~\cite{Blais04}, dispersive readout using 
a state-dependent shift $\chi$ of the cavity resonance is commonly 
used~\cite{Wallraff05, Mallet09}. 
In this scheme, 
it is important to choose a proper value of the cavity decay rate $\kappa$ in order to maximize the signal-to-noise ratio. 
Large $\kappa$ is favorable in terms of the information flow rate, 
but unfavorable in terms of the state distinguishability. 
As shown in Ref.~\onlinecite{Gambetta08}, under a fixed number of photons in the resonator, 
the signal-to-noise ratio is optimized by $\kappa=2\chi$ for a given $\chi$. 
Thus, large $\chi$ is advantageous because it allows large optimized $\kappa$. 
Achieving large $\chi$ is also important for applications 
such as conditional phase gate~\cite{DiCarlo09} and photon number resolving~\cite{Schuster07}. 

The magnitude of $\chi$ can be increased by making the coupling between 
the qubit and the cavity larger or decreasing the detuning between the cavity 
and the qubit, but both of them are at the price of reducing the upper limit 
of the qubit relaxation time through the cavity, so-called Purcell effect~\cite{Houck08}. 
Koch {\it et al.} pointed out another possibility of 
utilizing the third level of transmon qubits~\cite{Koch07}. 
The key requirement is to realize the condition $\omega_{01} < \omega_{\rm r} <\omega_{12}$, 
where $\omega_{\rm r}$ is the cavity resonant frequency and 
$\omega_{ij}$ is the qubit transition frequency between $i$-th and $j$-th levels. Achieving this condition, which they call a ``{\it straddling regime}", enables the cooperative interplay between 0--1 and 1--2 transitions, giving rise to the enhancement of $\chi$ without sacrificing the energy relaxation time of the qubit. However, in order to be in the dispersive regime, it is also required 
that the coupling strength $g$ between the qubit and the cavity 
is much smaller than the qubit anharmonicity, $|\omega_{12} - \omega_{01}|$, 
which is only a few-hundred~MHz for transmons. 

In this article, we demonstrate the straddling regime in a superconducting flux qubit coupled to a coplanar waveguide (CPW) resonator, and 
obtain 2$\chi$ as large as 2$\pi \times$80~MHz with a detuning of 5~GHz 
at the optimal bias point of the qubit. 
This values of 2$\chi$ is more than four times larger than that estimated from a simple two-level approximation.
The large anharmonicity of the flux qubit allows $g/2\pi$ of a few-hundred~MHz 
while staying in the deep dispersive regime. 
The flux qubit is coupled to the resonator by a capacitance 
instead of an inductance typically used~\cite{Abdumalikov08}. 
We find that the flux qubits can be strongly coupled to the resonator 
by the capacitance~\cite{Steffen10}, and actually 
the capacitive coupling is important for the observed large $\chi$ 
because it produces large matrix element for 1--2 transition. 
The capacitive coupling is also advantageous 
in terms of the readout backaction, as demonstrated 
in a transmon~\cite{Mallet09}, another dc-charge-insensitive device. 
Using the large dispersive shift, we achieved 90$\%$ contrast of the Rabi oscillations which is mainly limited by the energy relaxation of the qubit. 

Figure~1(a) shows the circuit diagram of the device we measure. 
The flux qubit is a conventional three-Josephson-junction flux qubit, 
in which one junction is made smaller than the other two by a factor of $\alpha$. One of the two islands defining the smaller junction is coupled to the center conductor of the CPW resonator via the coupling capacitance $C_{\rm c}$, 
while the other is connected to the ground. 
The Hamiltonian of the system consists of the qubit part, 
the resonator part, and the coupling part, that is,
\begin{equation}\label{fullHami}
H=H_{\rm q}+H_{\rm r}+H_{\rm c}.
\end{equation}
Each term of the Hamiltonian is given as below~\cite{SuppMat}:
\begin{eqnarray}\label{QHami}
\nonumber
&H_{\rm q}&=4E_{\rm C} \left[\frac{V}{X}
(n_1^2 + n_2^2)+2\frac{W}{X}n_1n_2 \right]\\
&&-E_{\rm J}\left[ \cos{\delta_1}-\cos{\delta_2}-
\alpha \cos(\delta_1-\delta_2+2\pi f) \right],
\end{eqnarray}
\begin{equation}\label{RHami}
H_{\rm r}=E_{\rm r}\sqrt{\frac{Y}{X}}(a^\dagger a+\frac{1}{2}),
\end{equation}
\begin{equation}\label{CHami}
H_{\rm c}=-2i\frac{\sqrt{\beta\gamma}}{Y^{1/4}X^{3/4}}
\sqrt{E_{\rm r}E_{\rm C}}(n_1 - n_2)(a^\dagger - a),
\end{equation}
where $E_{\rm J}$=$I_0$$\Phi_0$/2$\pi$,
$\Phi_0$=$h$/2$e$,
$f=\Phi / \Phi_0$,
$E_{\rm C}$=$e^2$/2$C_{\rm J}$,
$E_{\rm r}$=$\hbar$/$\sqrt{L_{\rm r}C_{\rm r}}$,
$V=(1+\alpha)(1+\gamma)+\beta$,
$W=\alpha(1+\gamma)+\beta$,
$X=(1+2\alpha)(1+\gamma)+2\beta$,
$Y=1+2\alpha+2\beta$,
$\beta$=$C_{\rm c}$/$C_{\rm J}$,
and
$\gamma$=$C_{\rm c}$/$C_{\rm r}$.
Here, $\Phi$ is an external flux, and $I_0$ and $C_{\rm J}$ are the critical current and the capacitance of the larger Josephson junction of the qubit, respectively. 
$\delta_i$ and $n_i$ ($i$=1, 2) are the phase difference across the larger junction, and its conjugate variable representing the charge number, respectively. 
The annihilation (creation) operator of a photon in the resonator is denoted as $a$ $(a^\dagger)$. 
Finally, $L_{\rm r}$ and $C_{\rm r}$ are the equivalent inductance and capacitance of the resonator, respectively.

\begin{figure}
\includegraphics[width=8.5cm]{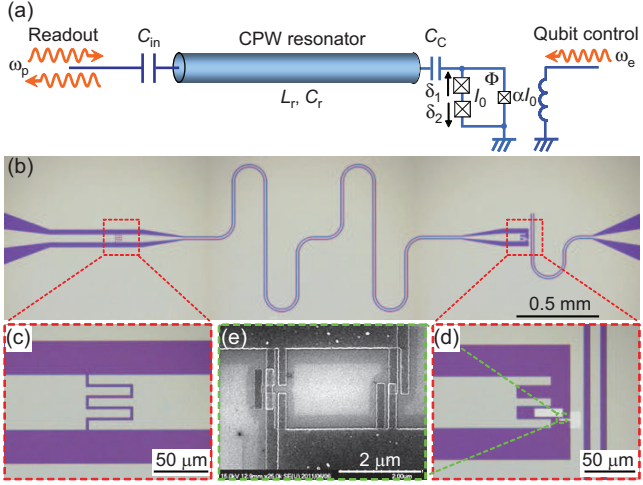}
\label{fig1}
\caption{(Color online) Circuit diagram and sample images.
(a) Circuit diagram of the device.
(b) Optical image of the device. 
The meandering trace in the center is a half-wavelength CPW resonator made of 50-nm-thick Nb.
(c) Interdigitated capacitance between the microwave feedline and the resonator, which is designed to be 15~fF. The gap of the capacitance is 2-$\mu$m wide.
(d) Magnified image of the qubit part. 
The qubit made of Al is located in the gap between the center conductor and the ground plane of the CPW resonator. 
One of the islands containing the smaller junction is made larger (a white rectangle in the center) so that it has large enough capacitance $C_{\rm c}$ to the center conductor of the resonator.
(e) Scanning electron micrograph of the three-junction flux qubit.
}
\end{figure}

Figure~1(b) shows an optical image of the device.
The half-wavelength CPW resonator is etched out of a 50-nm-thick Nb film sputtered on an oxidized high-resistivity silicon wafer. 
The fundamental-mode frequency $\omega_{\rm r}/2\pi$ is 10.656~GHz at $f=0.5$, and the $Q$ factor is 650, 
which is limited by the coupling to the feedline via the input capacitance $C_{\rm in}$ [Fig.~1(c)]~\cite{JJNote}.
The flux qubit [Fig.~1(e)] was fabricated by standard shadow evaporation technique, where two Al layers separated by Al$_2$O$_3$ are deposited from different angles. The qubit is located at one end of the resonator, where the electric field is the maximum. 
The coupling capacitance between the resonator and the qubit $C_{\rm c}$ is designed to be 4~fF. 
The qubit-control line couples to the flux qubit via a mutual inductance of $\sim$0.1~pH. 
The sample is mounted on a dilution refrigerator and cooled to $\sim$10~mK. 

\begin{figure}
\includegraphics[width=8.5cm]{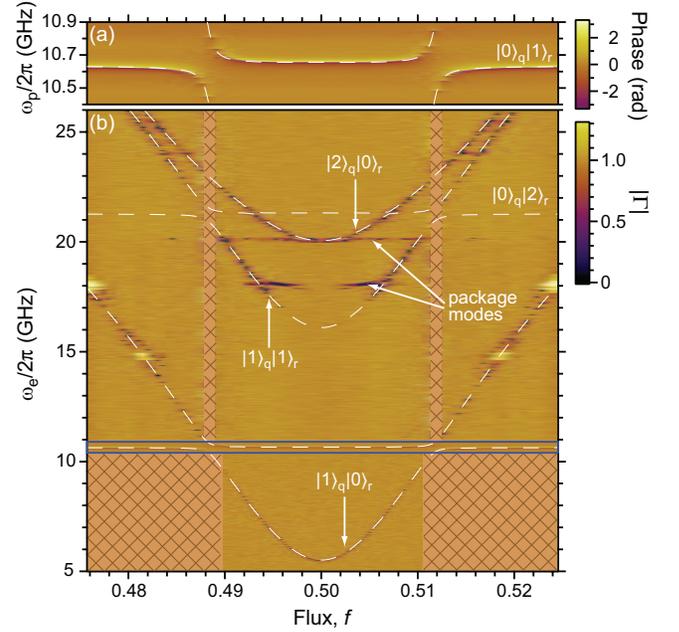}
\label{fig2}
\caption{(Color online) Spectroscopy of the coupled system. 
(a) Phase of the reflection coefficient $\Gamma$ as a function of the flux bias and the probe frequency. 
The vacuum Rabi splittings are clearly observed at $f=$ 0.488 and 0.512. 
(b) $\vert \Gamma \vert$ as a function of the flux bias and the qubit excitation frequency. 
$\vert \Gamma \vert$ in the panel is normalized by $\vert \Gamma \vert$ measured at $\omega_{\rm r}$.
Annotations on the energy levels denote the corresponding excitation from the ground state; 
$|i\rangle_{\rm q}|j\rangle_{\rm r}$ indicates that the qubit and the resonator are in the $i$-th and $j$-th states, respectively. 
The dashed curves represent the fit by Eq.~(\ref{fullHami}).
The same data as in (a) is shown in the blue box for comparison.
There is no data in the regions filled with the meshes.
}
\end{figure}

To confirm the coupling between the qubit and the resonator, 
we first measured the reflection coefficient $\Gamma$ of the resonator using a vector network analyzer. 
Figure~2(a) shows the phase of $\Gamma$ as a function of the flux bias and the probe frequency $\omega_{\rm p}/2\pi$. 
The power applied to the resonator is $-138$~dBm, corresponding to 0.08 photons in the resonator. 
Vacuum Rabi splittings are clearly observed at the points 
where the qubit energy $\hbar$$\omega_{01}$ is equal to $\hbar$$\omega_{\rm r}$, indicating that the strong coupling regime is achieved. The observed splitting size is 460~MHz. 

In order to measure the energy levels of the qubit in the wider frequency range, we applied another microwave at the frequency of $\omega_{\rm e}/2\pi$ 
to the control port continuously. 
We measure $\vert \Gamma \vert$ at fixed $\omega_{\rm p}$ of $2\pi \times$10.656~GHz, while sweeping $\omega_{\rm e}$ 
and the flux bias. Changes in $\vert \Gamma \vert$ is observed when the qubit is excited, thus revealing the energy band structure. 
Figure~2(b) shows $\vert \Gamma \vert$ as a function of $\omega_{\rm e}$ and the flux bias. 
As shown in the figure, each of the observed energy levels can be assigned to 
a particular excitation of the coupled system from the ground state. 
The resonances observed at around 18 and 20~GHz are probably due to resonant modes in the sample package. 
By fitting the data to the Hamiltonian [Eq.~(1)], we obtained the following circuit parameters: 
$E_{\rm J}$/$h$=148.4~GHz, $E_{\rm C}$/$h$=3.268~GHz, $\alpha$=0.6106, and $C_{\rm c}$=4.079~fF.
All of these values are quite consistent with our design. 
From these parameters, 0--1 and 1--2 transition frequencies of the qubit 
at $f=0.5$ are calculated to be $\omega_{01}/2\pi=$5.513~GHz and 
$\omega_{12}/2\pi=$14.54~GHz, respectively. 
Note that $\omega_{\rm r}$ is almost right at the middle of those frequencies. 
We come back to this point later. 

Next, we measured the dispersive shift of the cavity resonance depending on the qubit states by using pulsed microwaves and heterodyne detection of the reflected probe signal [Fig.~3(b) inset; see also Ref.~11]. The qubit is biased at the optimal point ($f=0.5$) where the dephasing due to the flux noise is minimal. When the qubit is prepared in the ground state, i.e., for the case without a qubit excitation pulse, we observe the cavity resonance at 10.656~GHz. It is seen in Fig.~3(a) as a dip in the amplitude of the normalized reflection coefficient $\Gamma' \equiv (\Gamma - \Gamma_{\rm on})/\vert \Gamma_{\rm off} -  \Gamma_{\rm on}\vert$, where $\Gamma_{\rm on}$ and $\Gamma_{\rm off}$ are the reflection coefficients obtained on- and off-resonance of the cavity, respectively. For the latter, the value of $\Gamma$ at $\omega_{\rm p}/2\pi=10.52$~GHz are used as a reference. When we apply a $\pi$-pulse to the qubit, on the other hand, a new dip appears at $\omega_{\rm p}/2\pi = 10.576$~GHz, while the one at 10.656~GHz is diminished significantly.
The magnitude of the observed dispersive shift (2$\vert \chi \vert$) is $2\pi \times$80~MHz.
Since the dispersive shift corresponds to the difference in the single-photon excitation energy of the resonator depending on the qubit states $|0\rangle$ or $|1\rangle$, we can estimate $2\vert \chi \vert$ from the energy band calculation using the circuit parameters obtained from the fitting. 
The calculated $2\vert \chi \vert$ is $2\pi \times$71.5~MHz~\cite{hlNote}, 
which is close to our observation. 
Presently, we do not fully understand the reason for the $\sim$10$\%$ discrepancy, though we suspect that the nonlinearity of the resonator exerts some influence on the measured dispersive shift~\cite{JJNote}.
Note, however, that both the observed and the calculated $2\vert \chi \vert$ are much larger than that expected from the standard Jaynes-Cummings Hamiltonian, 
in which the qubit is assumed to be an ideal two-level system and $\chi = g^2/\Delta$, where $g$ and $\Delta$ are the coupling strength and the detuning between the qubit and the resonator, respectively. 
In the present case, $2g$ is 2$\pi \times$460~MHz [Fig.~2(a)] and $\Delta$ is $2\pi \times$5.143~GHz, which leads to $2\vert \chi \vert$ of $2\pi \times$20.6~MHz. This value becomes even smaller if we take into account  a contribution of the Bloch-Siegert shift, namely $2\vert \chi \vert=2g^2/\Delta-2g^2/(\omega_{\rm r}+\omega_{01})=2\pi\times$14~MHz~\cite{Forn10}.
These facts indicate non-negligible contribution from higher energy levels of the flux qubit. Below we elaborate this point. 

\begin{figure}
\includegraphics[width=8.5cm]{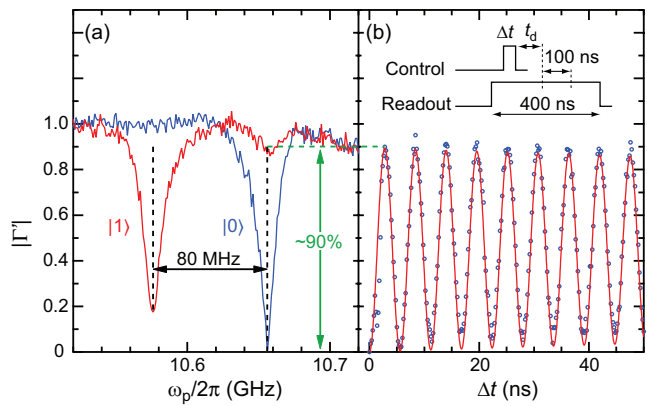}
\label{fig3}
\caption{(Color online) Dispersive shift and Rabi oscillations observed at $f$=0.5. 
(a) Amplitude of the normalized reflection coefficient $\Gamma'$ as a function of $\omega_{\rm p}$. 
The blue curve is taken when the qubit is in $\vert 0 \rangle$ state, 
and the red curve is when the qubit is in $\vert 1 \rangle$ state 
which is prepared by a $\pi$-pulse of 3-ns long. 
The shift of the dip frequency represents the dispersive shift. 
The residual dip at $\omega_{\rm p}/2\pi$=10.656~GHz for the qubit in
$\vert 1 \rangle$ state can be explained by the energy relaxation of the qubit. 
(b)~High contrast Rabi oscillations. 
$\vert \Gamma' \vert$ is plotted as a function of the control pulse length. 
The blue dots are the data and the red curve is a fit by a sinusoidal function with exponential decay. The inset depicts the applied pulse sequence. 
After the control pulse with the length of $\Delta t$ is applied, 
we wait for $t_{\rm d}$ before acquiring the readout signal. 
The length of the readout pulse was 400~ns, and the IF frequency was 50~MHz. The signal was sampled at 1~GS/s, and a 100-ns-long time trace was used to extract the amplitude and the phase.
}
\end{figure}

Following the work on transmon qubits~\cite{Koch07}, we consider the generalized Rabi model where we take into account the effect of the higher energy levels and counter-rotating terms.
The effective dispersive shift in the present system is given by 
\begin{equation}\label{DS}
\chi_{\rm eff} = \chi_{01} - \chi_{10} +
\frac{1}{2}\sum^{N}_{j=2}(\chi_{j1} - \chi_{1j} -\chi_{j0} + \chi_{0j}),
\end{equation}
where $\chi_{ij}=\vert g_{ij} \vert^2/(\omega_{ij}-\omega_{\rm r})$, 
$g_{ij}=\langle i \vert H_{\rm c} \vert j \rangle$, 
$\omega_{ij}$=$\omega_j-\omega_i$,
and $N$ is the number of qubit states taken into account. 
Here, $\hbar \omega_i$ represents the energy of the $i$-th qubit state. 
Unlike the case for a transmon, matrix elements $g_{ij}$ can be large not only for neighboring levels, i.e., $i = j\pm1$, but also for other pairs of levels because of large anharmonicity of the flux qubit.  
It is also worth mentioning that in the above formula, 
the contributions of the counter-rotating terms ($\chi_{ij}$ with $i>j$) are taken into account. 
As shown below, $\chi_{21}$ has a non-negligible 
contribution to $\chi_{\rm eff}$. 

Based on the parameters obtained from the fitting, we calculate the matrix elements $g_{ij}$ as a function of the flux bias~[Fig.~4(a)]. 
Remarkably, $|g_{12}|$ is more than two times larger than $|g_{01}|$ at $f$ = 0.5. In Fig.~4(b), we show the flux dependence of $\chi_{ij}$'s in Eq.~(5). 
Only the three main components ($\chi_{01}$, $\chi_{12}$, $\chi_{21}$) and $\chi_{\rm eff}$ with $N=4$ are plotted. 
Reflecting the large $|g_{12}|$ and the fact that $|\omega_{12}-\omega_{\rm r}| \sim |\omega_{01}-\omega_{\rm r}|$, $|\chi_{12}|$ is much larger than $|\chi_{01}|$ which corresponds to the dispersive shift estimated under the two-level approximation~\cite{effCcNote}. 
\begin{figure}[h]
\includegraphics[width=6.6cm]{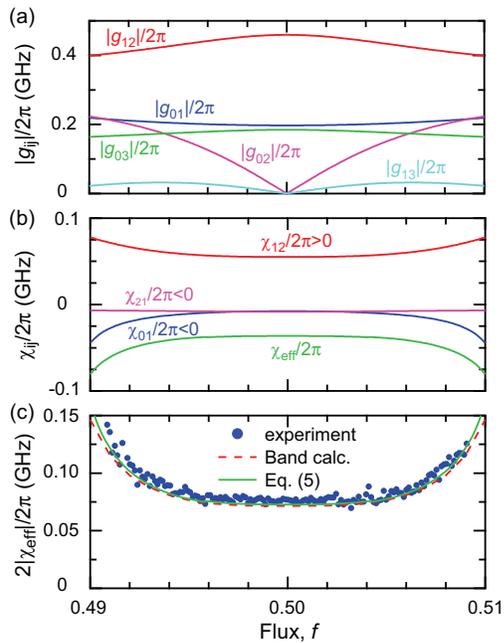}
\label{fig4}
\caption{(Color online) Flux bias dependence of $\vert g_{ij}\vert$, $\chi_{ij}$, and 2$\vert \chi_{\rm eff}\vert$.
(a) Calculated matrix elements of the coupling Hamiltonian $H_{\rm c}$ 
between the two states $\vert i\rangle$ and $\vert j\rangle$, 
i.e. $\vert g_{ij}\vert=\vert \langle i \vert H_{\rm c} \vert j \rangle \vert$
($i=0, 1$, and $0 \leq j \leq 3$).
(b) Calculated $\chi_{ij}$ and $\chi_{\rm eff}$. 
The three largest components are plotted here. 
The $\chi_{\rm eff}$ includes the terms up to $j=4$ in Eq.~(\ref{DS}).
(c) 2$\vert \chi_{\rm eff}\vert$ obtained from the measurement (blue dots), 
estimated from the energy band calculation (dashed red line), 
and calculated from Eq.~(\ref{DS}) (solid green line). 
Experimental $\chi_{\rm eff}$ was obtained by Lorentzian fit to the measured spectra, and has errors of a few MHz. }
\end{figure}
Moreover, because we are in the straddling regime, $\omega_{01} < \omega_{\rm r} < \omega_{12}$, 
all the three main components of $\chi_{ij}$'s sum up in a constructive way in Eq.~(5), which leads to the large $|\chi_{\rm eff}|$. 
At $f$=0.5, 2$\vert \chi_{\rm eff}\vert$ is calculated to be $2\pi \times$72.7~MHz, which is consistent with the band calculation and the experiment. 
If we use a formula for the dispersive shift under the two-level (and the rotating-wave) approximation, $\chi = g^2/\Delta$, and plug in the values of $g$ and $\chi$ from our experiments, we get $\Delta = 2\pi \times$ 1.323~GHz. 
The actual $\Delta$ is approximately four times larger than the above $\Delta$, which is due to the straddling effect. Since the Purcell decay time for the dispersive coupling is proportional to $\Delta^{2}$~\cite{Houck08}, the upper limit to the qubit decoherence time is improved by about 16 times.

To further confirm the validity of our theoretical model, we measured the dispersive shift as a function of the flux bias, which is shown in Fig.~4(c). 
In this measurement, we obtained $|\chi_{\rm eff}|$ in a different way from Fig.~3(a). 
We performed qubit spectroscopy as shown in Fig.~2(b), but with higher power of the qubit-control microwave. 
Due to the ac Stark shift, we observe side peaks around the main resonant peak of the qubit. 
Those peaks correspond to different photon numbers in the resonator~\cite{Schuster07}, and their separation is equal to $2|\chi_{\rm eff}|$. 
As shown in the figure, the agreement between the experiment and the theory is fairly good. 

By exploiting the large dispersive shift, we measured the Rabi oscillations using the 
pulse sequence shown in the inset of Fig.~3(b). 
After a control pulse with the length of $\Delta t$ is applied, 
we wait for $t_{\rm d}$ before we start acquiring the readout signal to calculate $\Gamma$. 
As shown in Fig.~3(b), we attained the contrast more than 90$\%$ after we experimentally optimized $t_{\rm d}$ to be 75~ns. 
We also estimate $t_{\rm d}$ from the response time for the voltage/current in the resonator $2\kappa^{-1} = 2Q/\omega_{\rm r}$
and the energy relaxation time $T_1$ of the qubit. 
We measured $T_1$ at $f$=0.5 to be 680~ns (data not shown).
The product $(1-\exp(-2t/\kappa))\cdot \exp(-t/T_1)$ shows its maximum at $t=$70~ns, which agrees well with the experimentally optimized $t_{\rm d}$.
Because $1 - \exp(-t_{\rm d}/T_1)$ amounts to nearly 0.1, 
we attribute the 10\% loss of the contrast to the energy relaxation of the qubit during $t_{\rm d}$. 

In conclusion, we demonstrated strong coupling between the flux qubit and the 
CPW resonator via a capacitor. 
We observed a large dispersive shift of the cavity resonance and 
quantitatively accounted for its magnitude. 
The key for the observed large shift is first of all the condition $\omega_{01} < \omega_{\rm r} < \omega_{12}$ to realize the straddling regime. The large matrix element $\vert g_{12}\vert$ and the reduced anharmonicity, although the qubit is still very anharmonic compared to transmons or phase qubits, further enhance its effect.
In addition, we have demonstrated the Rabi oscillations with more than 90\% contrast. 
Because the capacitive coupling for the flux qubit is advantageous in terms of the readout backaction, 
the present result is an important step toward the high-fidelity single-shot readout of the flux qubit using Josephson bifurcation~\cite{Mallet09, Maxime12} or parametric amplifiers~\cite{Vijay11}. 

\vspace{0.5 cm}
We would like to thank Y. Kitagawa for preparing sputtered Nb wafers, O. Astafiev for discussions, and M. Neeley and D. Sank for technical advices.
This work was supported by the MEXT Kakenhi ``Quantum Cybernetics", the JSPS through its FIRST Program, and the NICT Commissioned Research.

\appendix

\section{Derivation of the full Hamiltonian}
Figure~5 shows the equivalent circuit diagram of the device we measured. 
We assume that the larger Josephson junctions have capacitance $C_{\rm J}$ each, and the smaller one has $\alpha C_{\rm J}$.
In the diagram, $L_{\rm r}$, $C_{\rm r}$, and $C_{\rm c}$ represent the equivalent inductance and capacitance of the resonator, and the coupling capacitance between the resonator and the flux qubit, respectively. The phase difference across the larger Josephson junctions and the inductance $L_{\rm r}$ are denoted as $\delta_{i}$ ($i$=1, 2) and $\delta_{\rm r}$, respectively. 
The Lagrangian of the system is given by 
\begin{widetext}
\begin{eqnarray}\nonumber
\mathcal{L} &=& \frac{C_{\rm J}}{2}\Bigl(\frac{\Phi_0}{2\pi}\Bigr)^2\dot{\delta_1}^2 + 
\frac{C_{\rm J}}{2}\Bigl(\frac{\Phi_0}{2\pi}\Bigr)^2\dot{\delta_2}^2 + 
\frac{\alpha C_{\rm J}}{2}\Bigl(\frac{\Phi_0}{2\pi}\Bigr)^2(\dot{\delta_1}-\dot{\delta_2})^2 \\
\nonumber
&&+E_{\rm J}\cos{\delta_1} + E_{\rm J}\cos{\delta_2} + \alpha E_{\rm J}\cos(\delta_1-\delta_2+2\pi f) \\
&&+\frac{C_{\rm c}}{2}\Bigl(\frac{\Phi_0}{2\pi}\Bigr)^2(\dot{\delta_{\rm r}}+\dot{\delta_1}-\dot{\delta_2})^2 + 
\frac{C_{\rm r}}{2}\Bigl(\frac{\Phi_0}{2\pi}\Bigr)^2\dot{\delta_{\rm r}}^2 + 
\frac{1}{2L_{\rm r}}\Bigl(\frac{\Phi_0}{2\pi}\Bigr)^2\delta_{\rm r}^2\, ,
\end{eqnarray}
\end{widetext}
where $E_{\rm J}=I_0 \Phi_0/2 \pi$ and $f=\Phi_{\rm ex}/\Phi_0$. 
This can be rewritten as 
\begin{widetext}
\begin{eqnarray}\nonumber
\mathcal{L} &=& \frac{1}{2}m_1'\dot{\delta_1}^2 + \frac{1}{2}m_2'\dot{\delta_2}^2 
- m_3\dot{\delta_1}\dot{\delta_2} + E_{\rm J}\cos{\delta_1} + E_{\rm J}\cos{\delta_2} + \alpha E_{\rm J}\cos(\delta_1-\delta_2+2\pi f) \\ \label{Lre}
&& - m_{\rm c}\dot{\delta_2}\dot{\delta_{\rm r}} + m_{\rm c}\dot{\delta_1}\dot{\delta_{\rm r}} 
+ \frac{1}{2}m_{\rm r}'\dot{\delta_{\rm r}}^2 + \frac{1}{2L_{\rm r}}\Bigl(\frac{\Phi_0}{2\pi}\Bigr)^2\delta_{\rm r}^2\, ,
\end{eqnarray}
\end{widetext}
where 
\begin{eqnarray}
m_1' &=& \Bigl(\frac{\Phi_0}{2\pi}\Bigr)^2[(1 + \alpha)C_{\rm J}+C_{\rm c}] \\
m_2' &=& \Bigl(\frac{\Phi_0}{2\pi}\Bigr)^2[(1 + \alpha)C_{\rm J}+C_{\rm c}] \\
m_3' &=& \Bigl(\frac{\Phi_0}{2\pi}\Bigr)^2(\alpha C_{\rm J}+C_{\rm c}) \\
m_{\rm c} &=& \Bigl(\frac{\Phi_0}{2\pi}\Bigr)^2 C_{\rm c} \\
m_{\rm r}' &=& \Bigl(\frac{\Phi_0}{2\pi}\Bigr)^2 (C_{\rm c} + C_{\rm r}). 
\end{eqnarray}

From this Lagrangian, we calculate the Hamiltonian of the system. 
 We define the generalized momenta ($p_i = \partial \mathcal{L}/\partial \dot{\delta}_i$), 
\begin{equation}\label{gmom}
\left( 
\begin{array}{c}
p_1 \\
p_2 \\
p_{\rm r}
\end{array} 
\right) = 
\left( 
\begin{array}{ccc}
m_1' & -m_3' & m_{\rm c} \\
-m_3' & m_2' & -m_{\rm c} \\
m_{\rm c} & -m_{\rm c} &m'_{\rm r}
\end{array} 
\right) 
\left( 
\begin{array}{c}
\dot{\delta}_1 \\
\dot{\delta}_2 \\
\dot{\delta}_{\rm r}
\end{array} 
\right), 
\end{equation}
\begin{equation}
\vec{p} \equiv \mathbf{M}\vec{\dot{\delta}}.
\end{equation}
By the Legendre transformation, $\mathcal{H}=\sum p_i\dot{\delta}_i-\mathcal{L}$, we obtain the Hamiltonian
\begin{equation}
\mathcal{H}=\frac{1}{2}\vec{p}^{~t}\mathbf{M^{-1}}\vec{p}-U\, ,
\end{equation}
where 
\begin{equation}
U=E_{\rm J}\cos\delta_1+E_{\rm J}\cos\delta_2+\alpha E_{\rm J}\cos(\delta_1-\delta_2+2\pi f)\, ,
\end{equation}
\begin{widetext}
\begin{equation}
\mathbf{M}^{-1}=\frac{1}{\det \mathbf{M}}\left(
\begin{array}{ccc}
m'_2m'_{\rm r}-m_c^2 & m'_3m'_{\rm r}-m_c^2 & (m'_3-m'_2)m_c \\
m'_3m'_{\rm r}-m_c^2 & m'_1m'_{\rm r}-m_c^2 & (m'_1-m'_3)m_c \\
(m'_3-m'_2)m_{\rm c} & (m'_1-m'_3)m_{\rm c} & m'_1m'_2-m_3^2
\end{array}
\right),
\end{equation}
\end{widetext}
and
\begin{equation}
\det(\mathbf{M})=(m_1'm_2'-m_3'^2)m_{\rm r}'+(2m_3'-m_1'-m_2')m_c^2.
\end{equation}
The Hamiltonian consists of three parts,
the qubit part $\mathcal{H}_{\rm q}$,
the resonator part $\mathcal{H}_{\rm r}$,
and the coupling part $\mathcal{H}_{\rm c}$.
\begin{equation}~\label{totalHami}
\mathcal{H}=\mathcal{H}_{\rm q}+\mathcal{H}_{\rm r}+\mathcal{H}_{\rm c}.
\end{equation}
The qubit part is given by
\begin{equation}
\mathcal{H}_{\rm q}=\frac{p_1^2}{2m_1''}+\frac{p_2^2}{2m_2''}
+\frac{p_1p_2}{m_3''}-U\, ,
\end{equation}
where
\begin{eqnarray}\nonumber
\frac{1}{m_1''}&=&\frac{1}{m_2''}
=\frac{m_2'm_{\rm r}'-m_{\rm c}^2}{\det(\mathbf{M})} \\
&=&\left(\frac{2\pi}{\Phi_0}\right)^2\frac{(1+\alpha)(1+\gamma)+\beta}{(1+2\alpha)(1+\gamma)+2\beta}\cdot\frac{1}{C_{\rm J}}\, , \\
\nonumber \\ \nonumber
\frac{1}{m_3''}&=&\frac{m_3'm_{\rm r}'-m_{\rm c}^2}{\det(\mathbf{M})} \\
&=&\left(\frac{2\pi}{\Phi_0}\right)^2\frac{\alpha(1+\gamma)+\beta}{(1+2\alpha)(1+\gamma)+2\beta}\cdot\frac{1}{C_{\rm J}}\, ,
\end{eqnarray}
and
\begin{eqnarray}
\beta&=&C_{\rm c}/C_{\rm J}\, ,\\
\gamma&=&C_{\rm c}/C_{\rm r}.
\end{eqnarray}
By using
\begin{eqnarray}
n_i&=&\frac{2\pi}{\Phi_0}\frac{p_i}{2e}\, ,\\ \label{chargE}
E_{\rm C}&=&\frac{e^2}{2C_{\rm J}}\, ,
\end{eqnarray}
$\mathcal{H}_{\rm q}$ is written as
\begin{widetext}
\begin{equation}
\mathcal{H}_{\rm q}=4E_{\rm C}
\frac{(1+\alpha)(1+\gamma)+\beta}{(1+2\alpha)(1+\gamma)+2\beta}(n_1^2+n_2^2)
+8E_{\rm C}\frac{\alpha(1+\gamma)+\beta}{(1+2\alpha)(1+\gamma)+2\beta}n_1n_2-U.
\end{equation}
\end{widetext}

The resonator part is given by
\begin{eqnarray}
\mathcal{H}_{\rm r}&=&\frac{p_{\rm r}^2}{2m_{\rm r}''}
+\frac{1}{2L_{\rm r}}\left( \frac{\Phi_0}{2\pi} \right)^2\delta_{\rm r}^2 \\
&\equiv& \frac{p_{\rm r}^2}{2m_{\rm r}''}
+\frac{1}{2}m_{\rm r}''\omega_{\rm r}^2\delta_{\rm r}^2\, ,
\end{eqnarray}
where
\begin{eqnarray}\nonumber
\frac{1}{m_{\rm r}''}&=&\frac{m_1'm_2'-m_3'^2}{\det(\mathbf{M})} \\
&=&\left(\frac{2\pi}{\Phi_0}\right)^2\frac{1+2\alpha+2\beta}{(1+2\alpha)(1+\gamma)+2\beta}\cdot \frac{1}{C_{\rm r}}.
\end{eqnarray}
From this, we get the loaded resonant frequency $\omega_{\rm r}$,
\begin{equation}
\omega_{\rm r}
=\frac{1}{\sqrt{L_{\rm r}C_{\rm r}}}\sqrt{\frac{1+2\alpha+2\beta}{(1+2\alpha)(1+\gamma)+2\beta}}.
\end{equation}
Using creation and annihilation operators, we write the resonator part of the Hamiltonian as
\begin{equation}
\mathcal{H}_{\rm r} = \hbar \omega_{\rm r} (a^\dagger a+\frac{1}{2})\, ,
\end{equation}
where
\begin{eqnarray}~\label{xquantum}
\delta_{\rm r} &=& 
\sqrt{\frac{\hbar}{2m_{\rm r}''\omega_{\rm r}}}(a^\dagger + a)\, ,\\ \label{pquant}
p_{\rm r} &=& i\sqrt{\frac{m_{\rm r}''\omega_{\rm r} \hbar}{2}}(a^\dagger - a).
\end{eqnarray}
Finally, the coupling part is given by
\begin{equation}
\mathcal{H}_{\rm c}=\frac{1}{m_{\rm c}'}(p_1-p_2)p_{\rm r}\, ,
\end{equation}
where
\begin{eqnarray}\nonumber
\frac{1}{m_{\rm c}'}&=&\frac{(m_3'-m_2')m_{\rm c}}{\det(\mathbf{M})} \\
&=&\left(\frac{2\pi}{\Phi_0}\right)^2\frac{-\gamma}{(1+2\alpha)(1+\gamma)+2\beta}\cdot \frac{1}{C_{\rm J}}.
\end{eqnarray}
By using Eqs.~\ref{chargE} and \ref{pquant}, $\mathcal{H}_{\rm c}$ can be transformed as
\begin{widetext}
\begin{equation}
\mathcal{H}_{\rm c}=\frac{-2i}{(1+2\alpha+2\beta)^{1/4}}
\sqrt{\frac{\beta \gamma}{[(1+2\alpha)(1+\gamma)+2\beta]^{3/2}}}
\sqrt{E_{\rm r}E_{\rm c}}(n_1-n_2)(a^{\dagger}-a).
\end{equation}
\end{widetext}

\section{Method}
\subsection{Experimental setup}
Figure~6 shows an experimental setup diagram for our measurements. 
Qubit control and readout microwave pulses are generated by mixing the continuous microwave with pulses of 50~MHz IF frequency generated by DACs developed by Martinis group at UCSB~\cite{DACNote}.
The pulses are sent through the input microwave lines each of which include attenuators of 42~dB in total. For the readout line, the microwave pulses are further attenuated by 20~dB, and are routed to the resonator through a circulator to separate the input and output (reflected) waves. 

The output line for readout includes a band-pass filter (9--11~GHz), a low-pass filter ($f_{\rm c}$=12.4~GHz), and three isolators with $\sim$20~dB isolation each (9--11~GHz) in order to prevent noise/blackbody radiation coming from the cryogenic HEMT amplifier. The reflected signal is amplified by the amplifier and a room-temperature amplifier with a total gain of $\sim$66~dB,
and mixed with a local oscillator at an I/Q mixer down to the IF frequency. The I and Q quadratures are sampled at 1~GS/s
by a digitizer. The data is averaged 6.5$\times10^4$ times
for each measurement point.

For the frequency-domain measurement (Fig.~2), we replace the readout and the control circuits with a vector network analyzer and a continuous microwave generator, respectively.

The excitation energies of the flux qubit are controlled by
the local flux bias $\Phi$ which is generated by
a superconducting coil wound around the sample package mounted inside the cryogenic magnetic shield.

\subsection{Spectroscopy data fitting}
Figure~7(a) and (b) show the fitting results of the measured spectra (the same as Fig.~2(a) and (b)) with the energy levels obtained from the Hamiltonian (Eq.~\ref{totalHami}). The spectra are well fitted using 11 charge states for each islands of the flux qubit and the number of levels for the resonator is truncated by 5. The following circuit parameters are extracted by the fitting of the energy levels:
$E_{\rm J}$/$h$=148.4~GHz, $E_{\rm C}$/$h$=3.268~GHz, $\alpha$=0.6106, and $C_{\rm c}$=4.079~fF. 
For comparison, the same data as in (a) and (b) are shown in (c) and (d), respectively.

\subsection{Normalization of measured signals}
Figure~8(a) shows the reflection coefficient $\Gamma$ measured with the pulsed microwave and the heterodyne setup shown in Fig.~6. The data is represented in the polar coordinates for the frequency range from 10.50 to 10.75~GHz. The point A (B) indicated by the green (purple) arrow gives $\Gamma$ at $\omega_{\rm p}/2\pi=\omega_{\rm r}/2\pi=$10.656~GHz when the qubit is in $\vert 0\rangle$ ($\vert 1\rangle$) state. When we measure the Rabi oscillations, namely, fix $\omega_{\rm p}$ at $\omega_{\rm r}$ and vary the length of the qubit control pulse, observed $\Gamma$  goes back and forth between the point A and B. The dashed line in Fig.~8(a) represents a locus of the signal of the Rabi oscillations. Thus, the maximum amplitude of the Rabi oscillations is given by $\vert \overrightarrow{\strut{\rm AB}}\vert$. 
The point C indicated by the light blue arrow is $\Gamma$ at an off-resonance of $\omega_{\rm p}/2\pi$=10.52~GHz, which is the reference data used to normalize the measured data. We define the normalized reflection coefficient 
$\Gamma' \equiv (\Gamma-\Gamma_{\rm on})/\vert \Gamma_{\rm off}-\Gamma_{\rm on}\vert$, where $\Gamma_{\rm on}$ and $\Gamma_{\rm off}$ correspond to $\overrightarrow{\strut{\rm A}}$ and $\overrightarrow{\strut{\rm C}}$, respectively, and $\vert \Gamma_{\rm off}-\Gamma_{\rm on}\vert$ corresponds to the expected maximum amplitude of the Rabi oscillations.
Figure~8(b) depicts $\Gamma'$ represented in the polar coordinates. The points A, B, and C in (a) are displaced to ${\rm A}'$ (the origin), ${\rm B}'$, and ${\rm C}'$, respectively. 
$\vert \Gamma' \vert$ is equal to zero at $\omega_{\rm p}=\omega_{\rm r}$ when the qubit is in $\vert 0\rangle$ state, and is equal to $\vert \overrightarrow{\strut{\rm A'B'}} \vert$ corresponding to the normalized maximum amplitude of the Rabi oscillations $\sim$0.9 when a $\pi$-pulse is applied.

\bibliography{DS_KI_ref}

\providecommand{\noopsort}[1]{}\providecommand{\singleletter}[1]{#1}%
\begin{thebibliography}{18}%
\makeatletter
\providecommand \@ifxundefined [1]{%
 \@ifx{#1\undefined}
}%
\providecommand \@ifnum [1]{%
 \ifnum #1\expandafter \@firstoftwo
 \else \expandafter \@secondoftwo
 \fi
}%
\providecommand \@ifx [1]{%
 \ifx #1\expandafter \@firstoftwo
 \else \expandafter \@secondoftwo
 \fi
}%
\providecommand \natexlab [1]{#1}%
\providecommand \enquote  [1]{``#1''}%
\providecommand \bibnamefont  [1]{#1}%
\providecommand \bibfnamefont [1]{#1}%
\providecommand \citenamefont [1]{#1}%
\providecommand \href@noop [0]{\@secondoftwo}%
\providecommand \href [0]{\begingroup \@sanitize@url \@href}%
\providecommand \@href[1]{\@@startlink{#1}\@@href}%
\providecommand \@@href[1]{\endgroup#1\@@endlink}%
\providecommand \@sanitize@url [0]{\catcode `\\12\catcode `\$12\catcode
  `\&12\catcode `\#12\catcode `\^12\catcode `\_12\catcode `\%12\relax}%
\providecommand \@@startlink[1]{}%
\providecommand \@@endlink[0]{}%
\providecommand \url  [0]{\begingroup\@sanitize@url \@url }%
\providecommand \@url [1]{\endgroup\@href {#1}{\urlprefix }}%
\providecommand \urlprefix  [0]{URL }%
\providecommand \Eprint [0]{\href }%
\providecommand \doibase [0]{http://dx.doi.org/}%
\providecommand \selectlanguage [0]{\@gobble}%
\providecommand \bibinfo  [0]{\@secondoftwo}%
\providecommand \bibfield  [0]{\@secondoftwo}%
\providecommand \translation [1]{[#1]}%
\providecommand \BibitemOpen [0]{}%
\providecommand \bibitemStop [0]{}%
\providecommand \bibitemNoStop [0]{.\EOS\space}%
\providecommand \EOS [0]{\spacefactor3000\relax}%
\providecommand \BibitemShut  [1]{\csname bibitem#1\endcsname}%
\let\auto@bib@innerbib\@empty
\bibitem [{\citenamefont {Blais}\ \emph {et~al.}(2004)\citenamefont {Blais},
  \citenamefont {Huang}, \citenamefont {Wallraff}, \citenamefont {Girvin},\
  and\ \citenamefont {Schoelkopf}}]{Blais04}%
  \BibitemOpen
  \bibfield  {author} {\bibinfo {author} {\bibfnamefont {A.}~\bibnamefont
  {Blais}}, \bibinfo {author} {\bibfnamefont {R.-S.}\ \bibnamefont {Huang}},
  \bibinfo {author} {\bibfnamefont {A.}~\bibnamefont {Wallraff}}, \bibinfo
  {author} {\bibfnamefont {S.~M.}\ \bibnamefont {Girvin}}, \ and\ \bibinfo
  {author} {\bibfnamefont {R.~J.}\ \bibnamefont {Schoelkopf}},\ }\href@noop {}
  {\bibfield  {journal} {\bibinfo  {journal} {Phys.\ Rev.\ A}\ }\textbf
  {\bibinfo {volume} {69}},\ \bibinfo {pages} {062320} (\bibinfo {year}
  {2004})}\BibitemShut {NoStop}%
\bibitem [{\citenamefont {Wallraff}\ \emph {et~al.}(2005)\citenamefont
  {Wallraff}, \citenamefont {Schuster}, \citenamefont {Blais}, \citenamefont
  {Frunzio}, \citenamefont {Majer}, \citenamefont {Devoret}, \citenamefont
  {Girvin},\ and\ \citenamefont {Schoelkopf}}]{Wallraff05}%
  \BibitemOpen
  \bibfield  {author} {\bibinfo {author} {\bibfnamefont {A.}~\bibnamefont
  {Wallraff}}, \bibinfo {author} {\bibfnamefont {D.~I.}\ \bibnamefont
  {Schuster}}, \bibinfo {author} {\bibfnamefont {A.}~\bibnamefont {Blais}},
  \bibinfo {author} {\bibfnamefont {L.}~\bibnamefont {Frunzio}}, \bibinfo
  {author} {\bibfnamefont {J.}~\bibnamefont {Majer}}, \bibinfo {author}
  {\bibfnamefont {M.~H.}\ \bibnamefont {Devoret}}, \bibinfo {author}
  {\bibfnamefont {S.~M.}\ \bibnamefont {Girvin}}, \ and\ \bibinfo {author}
  {\bibfnamefont {R.~J.}\ \bibnamefont {Schoelkopf}},\ }\href@noop {}
  {\bibfield  {journal} {\bibinfo  {journal} {Phys.\ Rev.\ Lett.}\ }\textbf
  {\bibinfo {volume} {95}},\ \bibinfo {pages} {060501} (\bibinfo {year}
  {2005})}\BibitemShut {NoStop}%
\bibitem [{\citenamefont {Mallet}\ \emph {et~al.}(2009)\citenamefont {Mallet},
  \citenamefont {Ong}, \citenamefont {Palacios-Laloy}, \citenamefont {Nguyen},
  \citenamefont {Bertet}, \citenamefont {Vion},\ and\ \citenamefont
  {Esteve}}]{Mallet09}%
  \BibitemOpen
  \bibfield  {author} {\bibinfo {author} {\bibfnamefont {F.}~\bibnamefont
  {Mallet}}, \bibinfo {author} {\bibfnamefont {F.~R.}\ \bibnamefont {Ong}},
  \bibinfo {author} {\bibfnamefont {A.}~\bibnamefont {Palacios-Laloy}},
  \bibinfo {author} {\bibfnamefont {F.}~\bibnamefont {Nguyen}}, \bibinfo
  {author} {\bibfnamefont {P.}~\bibnamefont {Bertet}}, \bibinfo {author}
  {\bibfnamefont {D.}~\bibnamefont {Vion}}, \ and\ \bibinfo {author}
  {\bibfnamefont {D.}~\bibnamefont {Esteve}},\ }\href@noop {} {\bibfield
  {journal} {\bibinfo  {journal} {Nature Phys.}\ }\textbf {\bibinfo {volume}
  {5}},\ \bibinfo {pages} {791} (\bibinfo {year} {2009})}\BibitemShut {NoStop}%
\bibitem [{\citenamefont {Gambetta}\ \emph {et~al.}(2008)\citenamefont
  {Gambetta}, \citenamefont {Blais}, \citenamefont {Boissonneault},
  \citenamefont {Houck}, \citenamefont {Schuster},\ and\ \citenamefont
  {Girvin}}]{Gambetta08}%
  \BibitemOpen
  \bibfield  {author} {\bibinfo {author} {\bibfnamefont {J.}~\bibnamefont
  {Gambetta}}, \bibinfo {author} {\bibfnamefont {A.}~\bibnamefont {Blais}},
  \bibinfo {author} {\bibfnamefont {M.}~\bibnamefont {Boissonneault}}, \bibinfo
  {author} {\bibfnamefont {A.~A.}\ \bibnamefont {Houck}}, \bibinfo {author}
  {\bibfnamefont {D.~I.}\ \bibnamefont {Schuster}}, \ and\ \bibinfo {author}
  {\bibfnamefont {S.~M.}\ \bibnamefont {Girvin}},\ }\href@noop {} {\bibfield
  {journal} {\bibinfo  {journal} {Phys.\ Rev.\ A}\ }\textbf {\bibinfo {volume}
  {77}},\ \bibinfo {pages} {012112} (\bibinfo {year} {2008})}\BibitemShut
  {NoStop}%
\bibitem [{\citenamefont {DiCarlo}\ \emph {et~al.}(2009)\citenamefont
  {DiCarlo}, \citenamefont {Chow}, \citenamefont {Gambetta}, \citenamefont
  {Bishop}, \citenamefont {Johnson}, \citenamefont {Shuster}, \citenamefont
  {Majer}, \citenamefont {Blais}, \citenamefont {Frunzio}, \citenamefont
  {Girvin},\ and\ \citenamefont {Schoelkopf}}]{DiCarlo09}%
  \BibitemOpen
  \bibfield  {author} {\bibinfo {author} {\bibfnamefont {L.}~\bibnamefont
  {DiCarlo}}, \bibinfo {author} {\bibfnamefont {J.~M.}\ \bibnamefont {Chow}},
  \bibinfo {author} {\bibfnamefont {J.~M.}\ \bibnamefont {Gambetta}}, \bibinfo
  {author} {\bibfnamefont {L.~S.}\ \bibnamefont {Bishop}}, \bibinfo {author}
  {\bibfnamefont {B.~R.}\ \bibnamefont {Johnson}}, \bibinfo {author}
  {\bibfnamefont {D.~I.}\ \bibnamefont {Shuster}}, \bibinfo {author}
  {\bibfnamefont {J.}~\bibnamefont {Majer}}, \bibinfo {author} {\bibfnamefont
  {A.}~\bibnamefont {Blais}}, \bibinfo {author} {\bibfnamefont
  {L.}~\bibnamefont {Frunzio}}, \bibinfo {author} {\bibfnamefont {S.~M.}\
  \bibnamefont {Girvin}}, \ and\ \bibinfo {author} {\bibfnamefont {R.~J.}\
  \bibnamefont {Schoelkopf}},\ }\href@noop {} {\bibfield  {journal} {\bibinfo
  {journal} {Nature}\ }\textbf {\bibinfo {volume} {460}},\ \bibinfo {pages}
  {240} (\bibinfo {year} {2009})}\BibitemShut {NoStop}%
\bibitem [{\citenamefont {Schuster}\ \emph {et~al.}(2007)\citenamefont
  {Schuster}, \citenamefont {Houck}, \citenamefont {Schreier}, \citenamefont
  {Wallraff}, \citenamefont {Gambetta}, \citenamefont {Blais}, \citenamefont
  {Frunzio}, \citenamefont {Majer}, \citenamefont {Johnson}, \citenamefont
  {Devoret}, \citenamefont {Girvin},\ and\ \citenamefont
  {Schoelkopf}}]{Schuster07}%
  \BibitemOpen
  \bibfield  {author} {\bibinfo {author} {\bibfnamefont {D.~I.}\ \bibnamefont
  {Schuster}}, \bibinfo {author} {\bibfnamefont {A.~A.}\ \bibnamefont {Houck}},
  \bibinfo {author} {\bibfnamefont {J.~A.}\ \bibnamefont {Schreier}}, \bibinfo
  {author} {\bibfnamefont {A.}~\bibnamefont {Wallraff}}, \bibinfo {author}
  {\bibfnamefont {J.~M.}\ \bibnamefont {Gambetta}}, \bibinfo {author}
  {\bibfnamefont {A.}~\bibnamefont {Blais}}, \bibinfo {author} {\bibfnamefont
  {L.}~\bibnamefont {Frunzio}}, \bibinfo {author} {\bibfnamefont
  {J.}~\bibnamefont {Majer}}, \bibinfo {author} {\bibfnamefont
  {B.}~\bibnamefont {Johnson}}, \bibinfo {author} {\bibfnamefont {M.~H.}\
  \bibnamefont {Devoret}}, \bibinfo {author} {\bibfnamefont {S.~M.}\
  \bibnamefont {Girvin}}, \ and\ \bibinfo {author} {\bibfnamefont {R.~J.}\
  \bibnamefont {Schoelkopf}},\ }\href@noop {} {\bibfield  {journal} {\bibinfo
  {journal} {Nature}\ }\textbf {\bibinfo {volume} {445}},\ \bibinfo {pages}
  {515} (\bibinfo {year} {2007})}\BibitemShut {NoStop}%
\bibitem [{\citenamefont {Houck}\ \emph {et~al.}(2008)\citenamefont {Houck},
  \citenamefont {Schreier}, \citenamefont {Johnson}, \citenamefont {Chow},
  \citenamefont {Koch}, \citenamefont {Gambetta}, \citenamefont {Schuster},
  \citenamefont {Frunzio}, \citenamefont {Devoret}, \citenamefont {Girvin},\
  and\ \citenamefont {Schoelkopf}}]{Houck08}%
  \BibitemOpen
  \bibfield  {author} {\bibinfo {author} {\bibfnamefont {A.~A.}\ \bibnamefont
  {Houck}}, \bibinfo {author} {\bibfnamefont {J.~A.}\ \bibnamefont {Schreier}},
  \bibinfo {author} {\bibfnamefont {B.~R.}\ \bibnamefont {Johnson}}, \bibinfo
  {author} {\bibfnamefont {J.~M.}\ \bibnamefont {Chow}}, \bibinfo {author}
  {\bibfnamefont {J.}~\bibnamefont {Koch}}, \bibinfo {author} {\bibfnamefont
  {J.~M.}\ \bibnamefont {Gambetta}}, \bibinfo {author} {\bibfnamefont {D.~I.}\
  \bibnamefont {Schuster}}, \bibinfo {author} {\bibfnamefont {L.}~\bibnamefont
  {Frunzio}}, \bibinfo {author} {\bibfnamefont {M.~H.}\ \bibnamefont
  {Devoret}}, \bibinfo {author} {\bibfnamefont {S.~M.}\ \bibnamefont {Girvin}},
  \ and\ \bibinfo {author} {\bibfnamefont {R.~J.}\ \bibnamefont {Schoelkopf}},\
  }\href@noop {} {\bibfield  {journal} {\bibinfo  {journal} {Phys.\ Rev.\
  Lett.}\ }\textbf {\bibinfo {volume} {101}},\ \bibinfo {pages} {080502}
  (\bibinfo {year} {2008})}\BibitemShut {NoStop}%
\bibitem [{\citenamefont {Koch}\ \emph {et~al.}(2007)\citenamefont {Koch},
  \citenamefont {Yu}, \citenamefont {Gambetta}, \citenamefont {Houck},
  \citenamefont {Schuster}, \citenamefont {Majer}, \citenamefont {Blais},
  \citenamefont {Devoret}, \citenamefont {Girvin},\ and\ \citenamefont
  {Schoelkopf}}]{Koch07}%
  \BibitemOpen
  \bibfield  {author} {\bibinfo {author} {\bibfnamefont {J.}~\bibnamefont
  {Koch}}, \bibinfo {author} {\bibfnamefont {T.~M.}\ \bibnamefont {Yu}},
  \bibinfo {author} {\bibfnamefont {J.}~\bibnamefont {Gambetta}}, \bibinfo
  {author} {\bibfnamefont {A.~A.}\ \bibnamefont {Houck}}, \bibinfo {author}
  {\bibfnamefont {D.~I.}\ \bibnamefont {Schuster}}, \bibinfo {author}
  {\bibfnamefont {J.}~\bibnamefont {Majer}}, \bibinfo {author} {\bibfnamefont
  {A.}~\bibnamefont {Blais}}, \bibinfo {author} {\bibfnamefont {M.~H.}\
  \bibnamefont {Devoret}}, \bibinfo {author} {\bibfnamefont {S.~M.}\
  \bibnamefont {Girvin}}, \ and\ \bibinfo {author} {\bibfnamefont {R.~J.}\
  \bibnamefont {Schoelkopf}},\ }\href@noop {} {\bibfield  {journal} {\bibinfo
  {journal} {Phys.\ Rev.\ A}\ }\textbf {\bibinfo {volume} {76}},\ \bibinfo
  {pages} {042319} (\bibinfo {year} {2007})}\BibitemShut {NoStop}%
\bibitem [{\citenamefont {{Abdumalikov,~Jr.}}\ \emph
  {et~al.}(2008)\citenamefont {{Abdumalikov,~Jr.}}, \citenamefont {Astafiev},
  \citenamefont {Nakamura}, \citenamefont {Pashkin},\ and\ \citenamefont
  {Tsai}}]{Abdumalikov08}%
  \BibitemOpen
  \bibfield  {author} {\bibinfo {author} {\bibfnamefont {A.~A.}\ \bibnamefont
  {{Abdumalikov,~Jr.}}}, \bibinfo {author} {\bibfnamefont {O.}~\bibnamefont
  {Astafiev}}, \bibinfo {author} {\bibfnamefont {Y.}~\bibnamefont {Nakamura}},
  \bibinfo {author} {\bibfnamefont {Y.~A.}\ \bibnamefont {Pashkin}}, \ and\
  \bibinfo {author} {\bibfnamefont {J.~S.}\ \bibnamefont {Tsai}},\ }\href@noop
  {} {\bibfield  {journal} {\bibinfo  {journal} {Phys.\ Rev.\ B}\ }\textbf
  {\bibinfo {volume} {78}},\ \bibinfo {pages} {180502} (\bibinfo {year}
  {2008})}\BibitemShut {NoStop}%
\bibitem [{\citenamefont {Steffen}\ \emph {et~al.}(2010)\citenamefont
  {Steffen}, \citenamefont {Kumar}, \citenamefont {DiVincenzo}, \citenamefont
  {Rozen}, \citenamefont {Keefe}, \citenamefont {Rothwell},\ and\ \citenamefont
  {Ketchen}}]{Steffen10}%
  \BibitemOpen
  \bibfield  {author} {\bibinfo {author} {\bibfnamefont {M.}~\bibnamefont
  {Steffen}}, \bibinfo {author} {\bibfnamefont {S.}~\bibnamefont {Kumar}},
  \bibinfo {author} {\bibfnamefont {D.~P.}\ \bibnamefont {DiVincenzo}},
  \bibinfo {author} {\bibfnamefont {J.~R.}\ \bibnamefont {Rozen}}, \bibinfo
  {author} {\bibfnamefont {G.~A.}\ \bibnamefont {Keefe}}, \bibinfo {author}
  {\bibfnamefont {M.~B.}\ \bibnamefont {Rothwell}}, \ and\ \bibinfo {author}
  {\bibfnamefont {M.~B.}\ \bibnamefont {Ketchen}},\ }\href@noop {} {\bibfield
  {journal} {\bibinfo  {journal} {Phys.\ Rev.\ Lett.}\ }\textbf {\bibinfo
  {volume} {105}},\ \bibinfo {pages} {100502} (\bibinfo {year}
  {2010})}\BibitemShut {NoStop}%
\bibitem [{Sup()}]{SuppMat}%
  \BibitemOpen
  \href@noop {} {}\bibinfo {note} {See Appendix}\BibitemShut {NoStop}%
\bibitem [{JJN()}]{JJNote}%
  \BibitemOpen
  \href@noop {} {}\bibinfo {note} {In fact, a Josephson junction with designed
  $I_{\rm C}=0.7~\mu$A is embedded at the middle of the center conductor of the
  resonator. In our experiments described in this article, however, we always
  operate the resonator in the linear regime, so that we neglect this junction,
  except when we calculate $L_{\rm r}$ and $C_{\rm r}$ from $E_{\rm
  r}$}\BibitemShut {NoStop}%
\bibitem [{hlN()}]{hlNote}%
  \BibitemOpen
  \href@noop {} {}\bibinfo {note} {Although we have not measured the dispersive
  shift for the $\vert 2 \rangle$ state of the qubit, the energy band
  calculation predicts it to be $-2\pi \times 28.2$~MHz (the same sign as that
  for the $\vert 1 \rangle$ state) at the optimal flux bias point. Thus, it
  should be possible to discriminate all the three lowest states of the
  qubit}\BibitemShut {NoStop}%
\bibitem [{\citenamefont {Forn-D\'{i}az}\ \emph {et~al.}(2010)\citenamefont
  {Forn-D\'{i}az}, \citenamefont {Lisenfeld}, \citenamefont {Marcos},
  \citenamefont {Garc\'{i}a-Ripoll}, \citenamefont {Solano}, \citenamefont
  {Harmans},\ and\ \citenamefont {Mooij}}]{Forn10}%
  \BibitemOpen
  \bibfield  {author} {\bibinfo {author} {\bibfnamefont {P.}~\bibnamefont
  {Forn-D\'{i}az}}, \bibinfo {author} {\bibfnamefont {J.}~\bibnamefont
  {Lisenfeld}}, \bibinfo {author} {\bibfnamefont {D.}~\bibnamefont {Marcos}},
  \bibinfo {author} {\bibfnamefont {J.~J.}\ \bibnamefont {Garc\'{i}a-Ripoll}},
  \bibinfo {author} {\bibfnamefont {E.}~\bibnamefont {Solano}}, \bibinfo
  {author} {\bibfnamefont {C.~J. P.~M.}\ \bibnamefont {Harmans}}, \ and\
  \bibinfo {author} {\bibfnamefont {J.~E.}\ \bibnamefont {Mooij}},\ }\href@noop
  {} {\bibfield  {journal} {\bibinfo  {journal} {Phys.\ Rev.\ Lett.}\ }\textbf
  {\bibinfo {volume} {105}},\ \bibinfo {pages} {237001} (\bibinfo {year}
  {2010})}\BibitemShut {NoStop}%
\bibitem [{eff()}]{effCcNote}%
  \BibitemOpen
  \href@noop {} {}\bibinfo {note} {We note that the capacitive coupling also
  reduces $\omega_{12}$ of the flux qubit and hence further enhances $\vert
  \chi_{12}\vert$. For the present values of $E_{\rm J}$ and $E_{\rm C}$,
  $\omega_{12}$ would be $2\pi \times 19.912$~GHz if $C_{\rm c}$ were zero,
  compared to the actual $\omega_{12}$ of $2\pi \times 14.540$~GHz}\BibitemShut
  {NoStop}%
\bibitem [{\citenamefont {Boissonneault}\ \emph {et~al.}(2012)\citenamefont
  {Boissonneault}, \citenamefont {Gambetta},\ and\ \citenamefont
  {Blais}}]{Maxime12}%
  \BibitemOpen
  \bibfield  {author} {\bibinfo {author} {\bibfnamefont {M.}~\bibnamefont
  {Boissonneault}}, \bibinfo {author} {\bibfnamefont {J.~M.}\ \bibnamefont
  {Gambetta}}, \ and\ \bibinfo {author} {\bibfnamefont {A.}~\bibnamefont
  {Blais}},\ }\href@noop {} {\bibfield  {journal} {\bibinfo  {journal} {Phys.\
  Rev.\ A}\ }\textbf {\bibinfo {volume} {86}},\ \bibinfo {pages} {022326}
  (\bibinfo {year} {2012})}\BibitemShut {NoStop}%
\bibitem [{\citenamefont {Vijay}\ \emph {et~al.}(2011)\citenamefont {Vijay},
  \citenamefont {Slichter},\ and\ \citenamefont {Siddiqi}}]{Vijay11}%
  \BibitemOpen
  \bibfield  {author} {\bibinfo {author} {\bibfnamefont {R.}~\bibnamefont
  {Vijay}}, \bibinfo {author} {\bibfnamefont {D.~H.}\ \bibnamefont {Slichter}},
  \ and\ \bibinfo {author} {\bibfnamefont {I.}~\bibnamefont {Siddiqi}},\
  }\href@noop {} {\bibfield  {journal} {\bibinfo  {journal} {Phys.\ Rev.\
  Lett.}\ }\textbf {\bibinfo {volume} {106}},\ \bibinfo {pages} {110502}
  (\bibinfo {year} {2011})}\BibitemShut {NoStop}%
\bibitem [{DAC()}]{DACNote}%
  \BibitemOpen
  \href@noop {} {}\bibinfo {note}
  {URL:~https://commando.physics.ucsb.edu/tw/view \\ /Electronics}\BibitemShut
  {NoStop}%
\end{thebibliography}%

\begin{figure*}
\includegraphics[width=4.3cm]{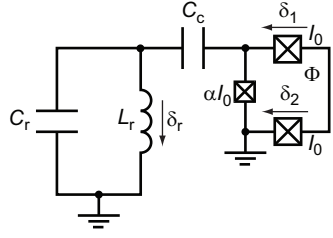}
\label{fig1s}
\caption{Circuit diagram of the flux qubit capacitively coupled to the $LC$ resonator.}
\end{figure*}

\begin{figure*}
\includegraphics[width=15cm]{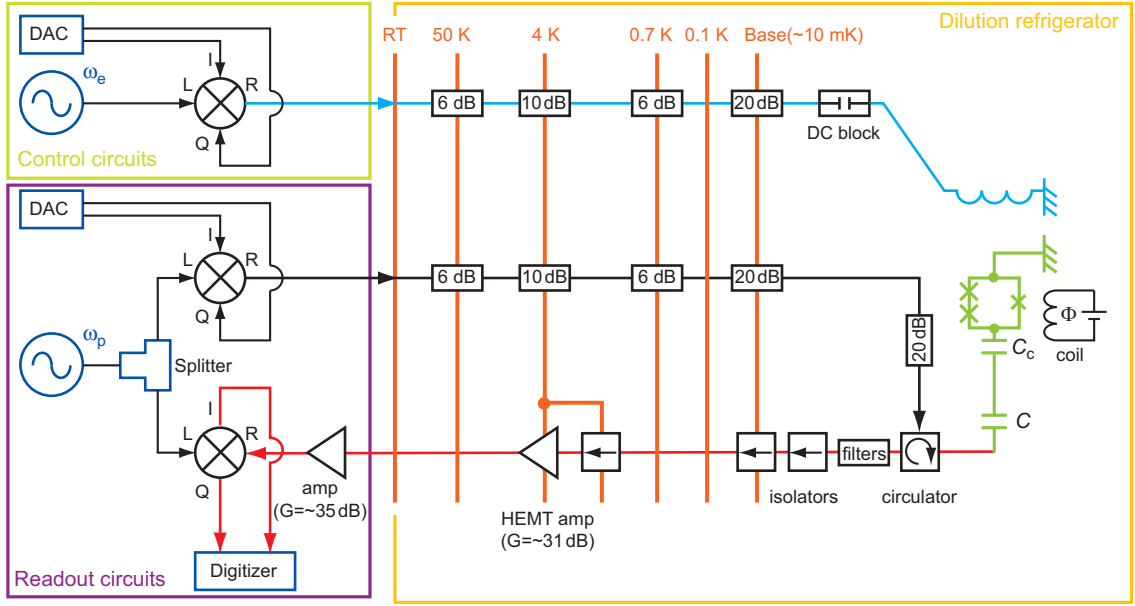}
\label{fig2s}
\caption{Experimental setup diagram for the heterodyne detection using the pulsed microwave.}
\end{figure*}

\begin{figure*}
\includegraphics[width=16.8cm]{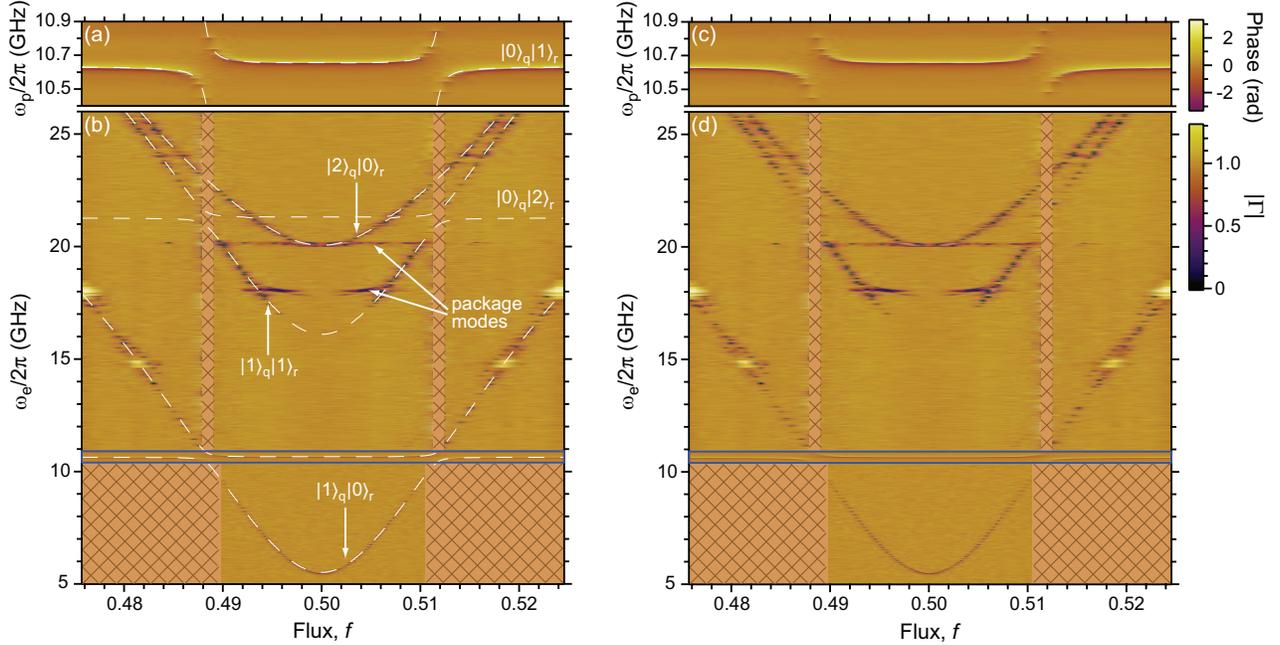}
\label{fig3s}
\caption{Spectra of the coupled system.
The dashed curves in (a) and (b) represent the fit to the energy levels calculated from the Hamiltonian, Eq.~(\ref{totalHami}). 
For comparison, the same data as in (a) and (b) are shown in (c) and (d), respectively.
The data in (a) [(c)] is embedded in (b) [(d)] in the blue box.
There is no data in the regions filled with the meshes.
}
\end{figure*}

\begin{figure*}
\includegraphics[width=16.8cm]{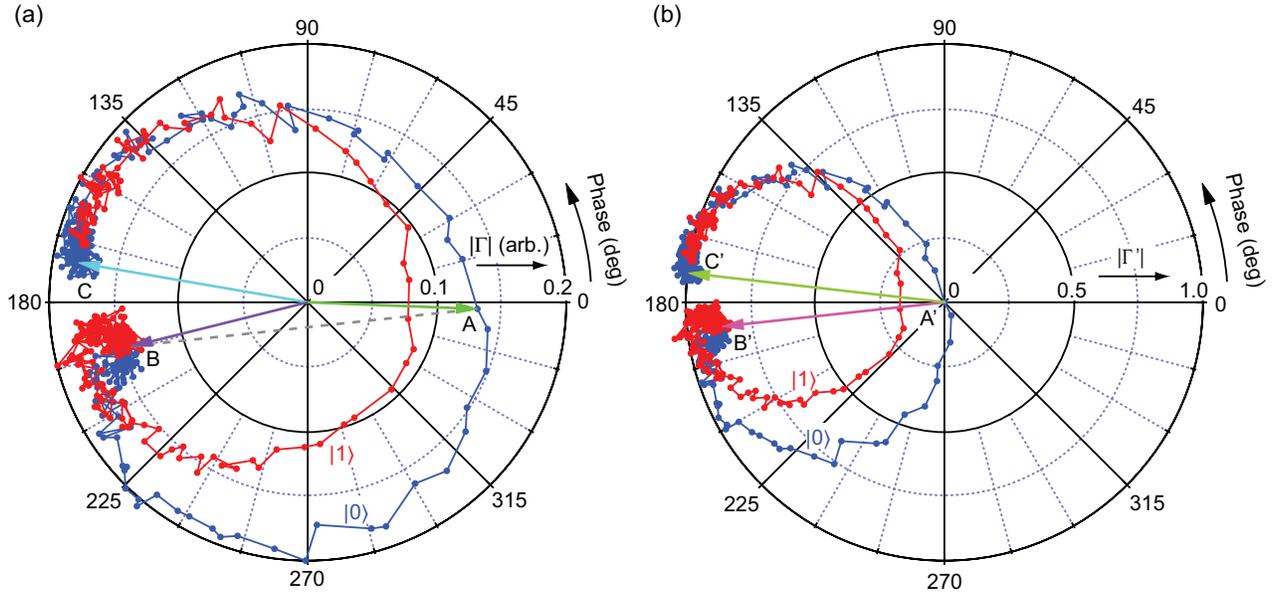}
\label{fig4s}
\caption[explain]{Normalization of the reflection coefficient $\Gamma$.
(a) $\Gamma$ represented in the polar coordinates for the frequency range from 10.50 to 10.75~GHz. The point ${\rm A}$ (${\rm B}$) corresponds to $\Gamma$ at $\omega_{\rm p}/2\pi=\omega_{\rm r}/2\pi$=10.656~GHz when the qubit is in $\vert 0\rangle$ ($\vert 1\rangle$) state.
The point C is $\Gamma$ at an off-resonance of $\omega_{\rm p}/2\pi=$10.52~GHz, which is a reference data used to normalize the measured data .
(b)~Normalized reflection coefficient $\Gamma'$ represented in the polar coordinates. The points A , B and C in (a) are displaced to ${\rm A}'$ (the origin), ${\rm B}'$, and ${\rm C}'$, respectively. The normalized maximum amplitude of the Rabi oscillation corresponds to $\vert \overrightarrow{\rm A'B'} \vert$ and it is $\sim$0.9.
}
\end{figure*}

\end{document}